\documentclass[a4paper,11pt]{revtex4}
\usepackage{amsfonts}
\usepackage{amssymb}
\usepackage{amsmath}

\def\QATOP#1#2{{#1 \atop #2}}
\begin{document}

\title{\mbox{Weyl correspondence method to construct multipartite entangled
quantum state }\thanks{%
Work supported by the President Foundation of Chinese Academy of Science and
the National Natural Science Foundation of China under grant 10475056}}
\author{Zhang Shuang-xi\thanks{%
shuangxi@mail.ustc.edu.cn}, Ma Xu  and Fan Hong-yi}
\affiliation{Department of Material Science and Engineering, University of Science and
Technology of China (USTC),\ Hefei, Anhui 230026, P.R. China}

\begin{abstract}
Via the Weyl correspondence approach, we construct multipartite entangled
state which is the common eigenvector of their center-of-mass coordinate and
mass-weighted relative momenta. This approach is concise and effective for
setting up the Fock representation of continuous multipartite entangled
states. The technique of integration within an ordered product (IWOP) of
operators is also essential in our derivation.
\end{abstract}

\maketitle

\section{Introduction}

\label{intro} By inventing the symbolic method, establishing quantum
mechanical representations and transformation theory in 1926, Dirac laid the
mathematical-physical foundation of quantum mechanics \cite{r1}.\ The
original work of comprising entanglement in quantum mechanics is introduced
in 1935 by Einstein, Podolsky, and Rosen, who formulated the EPR paradox, a
quantum-mechanical thought experiment designed to show that the theory is
incomplete. Now quantum entanglement, which is now considered as the feature
other than the hole of quantum mechanics, is widely understudied as a
physical resource, like energy, in quantum communication and quantum
information \cite{r2,r3,r4,r5,r6,r7,r8}. Introducing the entangled
representation will certainly help the study of entangled states. The EPR
state denoted as $\left\vert \eta \right\rangle $, as one of the simplest
bipartite entangled state is constructed in Ref. \cite{r9,r10,r11,r12},
which was enlightened by Einstein-Podolsky-Rosen's argument that two
particles' relative coordinate operator $Q_{1}-Q_{2}$ (we use capital \
letter and small letter to represent operator and number separately, and
henseforth) and the total momentum $P_{1}+P_{2}$ are commutable and can be
simultaneously measured \cite{r13}. It seems that introducing entangled
state representation was inevitable since many entangled problems can only
be clearly explained by it. An important question thus naturally arises: how
to concisely obtain the explicit form of multi-partite entangled state in
the Fock representations? Do we have a convenient approach for it? The
answer is affirmative. In\ this work with the help of method of integration
within an ordered product \ of operators (IWOP) \cite{r14,r15}, we shall
adopt Weyl correspondence (Weyl quantization scheme) to realize our goal,
and this approach is concise and effective for setting up the Fock
representation. To illustrate our approach clearly, in Sec. \ref{weyl} after
briefly introducing the Weyl correspondence rule, we demonstrate how the
Fock representation of bipartite entangled state $\left\vert \eta
\right\rangle $ can be derived via the Weyl correspondence approach. In Sec. %
\ref{bipartite}, \ref{tripartite} and \ref{multipartite} we discuss the
bipartite, tripartite case and multipartite EPR case respective. Via the
Weyl correspondence approach, we concentrate on the setting up the Fock
representation of continuous multipartite entangled states in this paper,
however, this is not to say the approach is limited to this scene.

\section{The Weyl correspondence approach and via which deriving bipartite
EPR state $\left\vert \protect\eta \right\rangle $}

\label{weyl}

Weyl correspondence is a quantization scheme which quantizes a classical
function $h\left( q,p\right) $ as an operator by the following integration%
\begin{equation}
H\left( P,Q\right) =\iint\limits_{-\infty }^{+\infty }dpdqh\left( p,q\right)
\Delta \left( q,p\right) ,  \label{1}
\end{equation}%
where $\Delta \left( q,p\right) $ is the Wigner operator \cite{r16}, its
original form in the coordinate representation is%
\begin{equation}
\Delta \left( q,p\right) =\frac{1}{2\pi }\int_{-\infty }^{+\infty
}dx^{\prime }e^{-ipx\prime }\left\vert q-\frac{x^{\prime }}{2}\right\rangle
\left\langle x+\frac{x^{\prime }}{2}\right\vert  \label{2}
\end{equation}%
In Ref. \cite{r17} we have performed this integral using $IWOP$ and get the
normally ordered form of $\Delta \left( q,p\right) ,$%
\begin{equation}
\Delta \left( q,p\right) =\frac{1}{\pi }{:}\exp \left[ -\left( q-Q\right)
^{2}-\left( p-P\right) ^{2}\right] \colon  \label{3}
\end{equation}%
where the symbol $:$ $:$ denotes normal ordering.

\bigskip In Ref. \cite{r19} Fan introduced the Weyl ordering by symbol ${{{{{%
{{{{{\QATOP{: }{:}}}}}}}}}}}$ ${{{{{{{{{{\QATOP{: }{:}}}}}}}}}}}$, which is
defined through the Weyl quantization scheme to quantize classical quantity
\cite{r18} $q^{m}p^{n}$ as
\begin{equation}
q^{m}p^{n}\Longrightarrow \left( \frac{1}{2}\right) ^{m}\sum_{l=0}^{m}\binom{%
m}{l}Q^{m-l}P^{n}Q^{l}\Longrightarrow {{{{{{{{{{\QATOP{: }{:}}}}}}}}}}}%
Q^{m}P^{n}{{{{{{{{{{\QATOP{: }{:}}}}}}}}}}},  \label{4}
\end{equation}%
and the Weyl ordering possesses three remarkable properties \cite{r19}:

(a)\ the order of Bose operators within a Weyl ordered product (or within
the Weyl ordering symbol ${{{{{{{{{{\QATOP{: }{:}}}}}}}}}}}$ ${{{{{{{{{{%
\QATOP{: }{:}}}}}}}}}}}$) can be permuted; the right-hand side of (\ref{4})
exhibits the definition of Weyl ordering, so%
\begin{eqnarray}
&&\left( \frac{1}{2}\right) ^{m}\sum_{l=0}^{m}\binom{m}{l}Q^{m-l}P^{n}Q^{l}
\notag \\
&=&{{{{{{{{{{\QATOP{: }{:}}}}}}}}}}}\left( \frac{1}{2}\right)
^{m}\sum_{l=0}^{m}\frac{m!}{l!\left( m-l\right) !}Q^{m-l}P^{n}Q^{l}{{{{{{{{{{%
\QATOP{: }{:}}}}}}}}}}}={{{{{{{{{{\QATOP{: }{:}}}}}}}}}}}Q^{m}P^{n}{{{{{{{{{{%
\QATOP{: }{:}}}}}}}}}}},
\end{eqnarray}%
which means
\begin{equation}
{{{{{{{{{{\QATOP{: }{:}}}}}}}}}}}Q^{m}P^{n}{{{{{{{{{{\QATOP{: }{:}}}}}}}}}}}%
=\iint\limits_{-\infty }^{\infty }\mathrm{d}p\mathrm{d}qq^{m}p^{n}\Delta
\left( q,p\right) .  \label{6}
\end{equation}%
It then follows the second property,

(b) Comparing (\ref{6}) with (\ref{1}) we derive the Weyl ordered form of
the Wigner operator is \cite{r19}
\begin{equation}
\triangle \left( q,p\right) ={{{{{{{{{{\QATOP{: }{:}}}}}}}}}}}\delta \left(
p-P\right) \delta \left( q-Q\right) {{{{{{{{{{\QATOP{: }{:}}}}}}}}}}}={{{{{{{%
{{{\QATOP{: }{:}}}}}}}}}}}\delta \left( q-Q\right) \delta \left( p-P\right) {%
{{{{{{{{{\QATOP{: }{:}}}}}}}}}}}.  \label{7}
\end{equation}%
Moreover, we can have the technique of integration within the Weyl ordered
product (IWWOP) of operators:

(c) a Weyl ordered product can be integrated with respect to a $c$-number
provided that the integration is convergent.

Thus the Weyl quantization rule for a classical function $h\left( p,q\right)
$ transiting to its quantum operator is
\begin{eqnarray}
&&\iint\limits_{-\infty }^{\infty }\mathrm{d}p\mathrm{d}qh\left( p,q\right)
\triangle \left( q,p\right)  \notag \\
&=&\iint\limits_{-\infty }^{\infty }\mathrm{d}p\mathrm{d}qh\left( p,q\right)
{{{{{{{{{{\QATOP{: }{:}}}}}}}}}}}\delta \left( p-P\right) \delta \left(
q-Q\right) {{{{{{{{{{\QATOP{: }{:}}}}}}}}}}}={{{{{{{{{{\QATOP{: }{:}}}}}}}}}}%
}h\left( P,Q\right) {{{{{{{{{{\QATOP{: }{:}}}}}}}}}}},
\end{eqnarray}%
which means that a Weyl ordered operator ${{{{{{{{{{\QATOP{: }{:}}}}}}}}}}}%
h\left( P,Q\right) {{{{{{{{{{\QATOP{: }{:}}}}}}}}}}}$'s classical
correspondence is just $h\left( p,q\right) .$

\bigskip $\left\vert \eta \right\rangle $ \cite{r9} is the bipartite EPR
state($\eta =\eta _{1}+\mathrm{i}\eta _{2}$), whose definite expression and
eigen-equations are

\begin{equation*}
\left\vert \eta \right\rangle =\exp [-|\eta |^{2}/2+\eta a_{1}^{\dagger
}-\eta ^{\ast }a_{2}^{\dagger }+a_{1}^{\dagger }a_{2}^{\dagger }]\left\vert
00\right\rangle
\end{equation*}

\begin{eqnarray*}
\left( Q_{1}-Q_{2}\right) \left\vert \eta \right\rangle &=&\sqrt{2}\eta
_{1}\left\vert \eta \right\rangle \\
\left( P_{1}-P_{2}\right) \left\vert \eta \right\rangle &=&\sqrt{2}\eta
_{2}\left\vert \eta \right\rangle
\end{eqnarray*}%
where in Fork space $a_{1}^{\dagger }$, $a_{2}^{\dagger }$ are generate
operators. The classical Weyl correspondence of the projector of the
projector operator $\left\vert \eta \right\rangle \left\langle \eta
\right\vert $ writes

\begin{equation}
\left\vert \eta \right\rangle \left\langle \eta \right\vert \Longrightarrow
\delta \left[ \sqrt{2}\eta _{1}-\left( q_{1}-q_{2}\right) \right] \delta %
\left[ \sqrt{2}\eta _{2}-\left( p_{1}+p_{2}\right) \right]
\end{equation}%
that is

\begin{eqnarray}
&&\left\vert \eta \right\rangle \left\langle \eta \right\vert  \notag \\
&\Longrightarrow &\iiiint \mathrm{d}q_{1}\mathrm{d}p_{1}\mathrm{d}q_{2}%
\mathrm{d}p_{2}\delta \left[ \sqrt{2}\eta _{1}-\left( q_{1}-q_{2}\right) %
\right] \delta \left[ \sqrt{2}\eta _{2}-\left( p_{1}+p_{2}\right) \right]
\Delta _{1}\left( q_{1},p_{1}\right) \Delta _{2}\left( q_{2},p_{2}\right)
\notag \\
&=&\iiiint \mathrm{d}q_{1}\mathrm{d}p_{1}\mathrm{d}q_{2}\mathrm{d}%
p_{2}\delta \left[ \sqrt{2}\eta _{1}-\left( q_{1}-q_{2}\right) \right]
\delta \left[ \sqrt{2}\eta _{2}-\left( p_{1}+p_{2}\right) \right]  \notag \\
&&\times \frac{1}{\pi ^{2}}\colon \exp \left[ -\left( q_{1}-Q_{1}\right)
^{2}-\left( p_{1}-P_{1}\right) ^{2}-\left( q_{2}-Q_{2}\right) ^{2}-\left(
p_{2}-P_{2}\right) ^{2}\right] \colon  \notag \\
&=&\iint \mathrm{d}q_{1}\mathrm{d}p_{1}\frac{1}{\pi ^{2}}\colon \exp \left[
-\left( q_{1}-Q_{1}\right) ^{2}-\left( p_{1}-P_{1}\right) ^{2}-\left( q_{1}-%
\sqrt{2}\eta _{1}-Q_{2}\right) ^{2}-\left( \sqrt{2}\eta
_{1}-p_{1}-P_{2}\right) ^{2}\right] \colon  \notag \\
&=&\frac{1}{2\pi }\colon \exp \left\{ \frac{1}{2}\left[ -\left[ \sqrt{2}\eta
_{1}-\left( Q_{1}-Q_{2}\right) \right] ^{2}-\left[ \sqrt{2}\eta _{2}-\left(
P_{1}+P_{2}\right) \right] ^{2}\right] \right\} \colon  \label{yita}
\end{eqnarray}

Due to%
\begin{equation}
Q_{i}=\left( a_{i}+a_{i}^{\dagger }\right) /\sqrt{2},\text{\qquad }%
P_{i}=\left( a_{i}-a_{i}^{\dagger }\right) /\left( \mathrm{i}\sqrt{2}\right)
,  \label{qp}
\end{equation}

where\ $\left[ a_{i},a_{j}^{\dagger }\right] =\delta _{ij}$, we rewrite (\ref%
{yita}) as

\begin{eqnarray*}
\left\vert \eta \right\rangle \left\langle \eta \right\vert &=&\left\vert
C\right\vert ^{2}\frac{1}{2\pi }\colon \exp \left\{ \frac{1}{2}\left[ -\left[
\sqrt{2}\eta _{1}-\left( Q_{1}-Q_{2}\right) \right] ^{2}-\left[ \sqrt{2}\eta
_{2}-\left( P_{1}+P_{2}\right) \right] ^{2}\right] \right\} \colon \\
&=&\left\vert C\right\vert ^{2}\frac{1}{2\pi }\colon \exp \left\{
-\left\vert \eta \right\vert ^{2}+\eta ^{\ast }a_{1}+\eta a_{1}^{\dagger
}-\eta a_{2}-\eta ^{\ast }a_{2}^{\dagger }-a_{1}^{\dagger
}a_{1}-a_{2}^{\dagger }a_{2}+a_{1}^{\dagger }a_{2}^{\dagger
}+a_{1}a_{2}\right\} \colon
\end{eqnarray*}

and by using $\colon \exp \left( -a_{1}^{\dag }a_{1}-a_{2}^{\dag
}a_{2}\right) \colon $ $=|00\rangle \langle 00|$, we can decompose the $%
\left\vert \eta \right\rangle \left\langle \eta \right\vert $ as following

\begin{eqnarray*}
&&\colon exp\left[ -\left\vert \eta \right\vert ^{2}+\eta ^{\ast }a_{1}+\eta
a_{1}^{\dagger }-\eta a_{2}-\eta ^{\ast }a_{2}^{\dagger }-a_{1}^{\dagger
}a_{1}-a_{2}^{\dagger }a_{2}+a_{1}^{\dagger }a_{2}^{\dagger }+a_{1}a_{2}%
\right] \colon \\
&=&exp\left[ -\left\vert \eta \right\vert ^{2}/2+\eta a_{1}^{\dagger }-\eta
^{\ast }a_{2}^{\dagger }+a_{1}^{\dagger }a_{2}^{\dagger }\right] \colon \exp %
\left[ -a_{1}^{\dagger }a_{1}-a_{2}^{\dagger }a_{2}\right] \colon \exp \left[
-\left\vert \eta \right\vert ^{2}/2+\eta ^{\ast }a_{1}-\eta a_{2}+a_{1}a_{2}%
\right] \\
&=&f(a_{1}^{\dagger },a_{2}^{\dagger })\left\vert 00\right\rangle
\left\langle 00\right\vert f^{\dagger }(a_{1}^{\dagger },a_{2}^{\dagger })
\end{eqnarray*}

and the normalized constant $C$ can be determined by normalized condition $%
\left\langle \eta \right\vert \left. \eta^{\prime } \right\rangle =\pi
\delta ^{(2)}\left( \eta -\eta ^{\prime }\right)$ up to random phase, so we
deduced the expression of $\left\vert \eta \right\rangle $ as

\begin{equation}
\left\vert \eta \right\rangle =exp\left[ -\left\vert \eta \right\vert
^{2}/2+\eta a_{1}^{\dagger }-\eta ^{\ast }a_{2}^{\dagger }+a_{1}^{\dagger
}a_{2}^{\dagger }\right] \left\vert 00\right\rangle  \label{state-yita}
\end{equation}

This is the bipartite EPR state, it is easy to check its (over)completeness
and orthogonal

\begin{eqnarray*}
\left\langle \eta \right\vert \left. \eta ^{\prime }\right\rangle &=&\pi
\delta ^{(2)}\left( \eta -\eta ^{\prime }\right) \\
\int \frac{\mathrm{d}^{2}\eta }{\pi }\left\vert \eta \right\rangle
\left\langle \eta \right\vert &=&1
\end{eqnarray*}

So far, we success deduce the expression of $\left\vert \eta \right\rangle $
by employing the Weyl correspondence approach. In the following sections, we
will also employ the same approach to derive the bipartite, tripartite and
multipartite representation of their center-of-mass coordinate and
mass-weighted relative momentum entangled state.

\section{Deriving the common eigenstates of bipartite's center-of-mass
cooridinate and mass-weighted relative momentum via Weyl correspondence
approach}

\label{bipartite}

In this section we will find the common eigenstates of bipartite's
center-of-mass coordinate and mass-weighted relative momentum through the
Weyl correspondence and the IWOP technique. Following our previous work, we
introduce the biparticle's center-of-mass coordinate operator $Q_{cm}=\mu
_{1}Q_{1}+\mu _{2}Q_{2},$ and the mass-weighted relative momentum operator $%
P_{r}=P_{1}/\mu _{1}-P_{2}/\mu _{2}$ of bipartite system, where $\mu _{i}$
is the relative of mass defined $\mu _{i}=m_{i}/\left( m_{1}+m_{2}\right) ,$
and $m_{i}$ is the mass of the $i$-th paritcle, and we have $\mu _{1}+\mu
_{2}=1$. It is easy to check that the two operators are compatible $\left[
Q_{cm},P_{r}\right] =0$, so it is meaningfull to construct the simultaneous
eigenstate of $Q_{cm}$ and $P_{r}$\ in terms of conventional creation and
annihilation operators, and we note its common eigenvector is $\left\vert
q_{cm},\varrho \right\rangle $, which is $\left\vert \xi \right\rangle $ in
Ref. \cite{r9} . In this section we shall employ the Weyl correspondence and
the IWOP technique to derive the explicit form of $\left\vert \xi
\right\rangle $ in two-mode Fock space, and our starting point are the the
eigen-equations write
\begin{subequations}
\begin{eqnarray}
\left( \mu _{1}Q_{1}+\mu _{2}Q_{2}\right) \left\vert q_{cm},\varrho
\right\rangle &=&q_{cm}\left\vert q_{cm},\varrho \right\rangle ,  \label{9}
\\
\text{ }\left( \frac{P_{1}}{\mu _{1}}-\frac{P_{2}}{\mu _{2}}\right)
\left\vert q_{cm},\varrho \right\rangle &=&\varrho \mathfrak{\ }\left\vert
q_{cm},\varrho \right\rangle ,
\end{eqnarray}

from which we can write the classical Weyl correspondence of the projector $%
\left\vert q_{cm},\varrho \right\rangle \left\langle q_{cm},\varrho
\right\vert $,

\end{subequations}
\begin{equation}
\left\vert q_{cm},\varrho \right\rangle \left\langle q_{cm},\varrho
\right\vert \Longrightarrow \delta \left[ q_{cm}-\left( \mu _{1}q_{1}+\mu
_{2}q_{2}\right) \right] \delta \left[ \varrho -\left( \frac{p_{1}}{\mu _{1}}%
-\frac{p_{2}}{\mu _{2}}\right) \right]  \label{10}
\end{equation}

According to (\ref{10}), we know the classical Weyl correspondence of the
projector $\left\vert q_{cm},\varrho \right\rangle \left\langle
q_{cm},\varrho \right\vert $ is

\begin{eqnarray}
&&\left\vert q_{cm},\varrho \right\rangle \left\langle q_{cm},\varrho
\right\vert  \notag \\
&\Longrightarrow &\iiiint \mathrm{d}q_{1}\mathrm{d}q_{2}\mathrm{d}p_{1}%
\mathrm{d}p_{2}\delta \left[ q_{cm}-\left( \mu _{1}q_{1}+\mu
_{2}q_{2}\right) \right] \delta \left[ \varrho -\left( \frac{p_{1}}{\mu _{1}}%
-\frac{p_{2}}{\mu _{2}}\right) \right] \Delta _{1}\left( q_{1},p_{1}\right)
\Delta _{2}\left( q_{2},p_{2}\right)  \notag \\
&=&\iiiint \mathrm{d}q_{1}\mathrm{d}q_{2}\mathrm{d}p_{1}\mathrm{d}%
p_{2}\delta \left[ q_{cm}-\left( \mu _{1}q_{1}+\mu _{2}q_{2}\right) \right]
\delta \left[ \varrho -\left( \frac{p_{1}}{\mu _{1}}-\frac{p_{2}}{\mu _{2}}%
\right) \right]  \notag \\
&&\times \frac{1}{\pi ^{2}}\colon \exp \left[ -\left( q_{1}-Q_{1}\right)
^{2}-\left( p_{1}-P_{1}\right) ^{2}-\left( q_{2}-Q_{2}\right) ^{2}-\left(
p_{2}-P_{2}\right) ^{2}\right] \colon  \notag \\
&=&\iint \mathrm{d}q_{1}\mathrm{d}p_{1}\frac{1}{\pi ^{2}}\colon \exp \left[
-\left( q_{1}-Q_{1}\right) ^{2}-\left( p_{1}-P_{1}\right) ^{2}-\left( \frac{%
q_{cm}-\mu _{1}q_{1}}{\mu _{2}}-Q_{2}\right) ^{2}-\left( \frac{\mu _{2}}{\mu
_{1}}p_{1}-\mu _{2}\varrho -P_{2}\right) ^{2}\right] \colon  \notag \\
&=&\frac{\mu _{1}\mu _{2}}{\pi \left( \mu _{1}^{2}+\mu _{2}^{2}\right) }%
\colon \exp \left\{ \frac{1}{\left( \mu _{1}^{2}+\mu _{2}^{2}\right) }\left[
-\left( q_{cm}-\left( \mu _{1}Q_{1}+\mu _{2}Q_{2}\right) \right) ^{2}-\mu
_{1}^{2}\mu _{2}^{2}\left( \varrho -\left( \frac{P_{1}}{\mu _{1}}-\frac{P_{2}%
}{\mu _{2}}\right) \right) ^{2}\right] \right\} \colon  \label{11}
\end{eqnarray}%
Due to (\ref{qp}) we rewrite (\ref{11}) as
\begin{equation}
\left\vert q_{cm},\varrho \right\rangle \left\langle q_{cm},\varrho
\right\vert =\left\vert C\right\vert ^{2}\frac{\mu _{1}\mu _{2}}{\pi \lambda
}\colon \exp \left\{ \frac{1}{\lambda }\left[ -\left( q_{cm}-\left( \mu
_{1}Q_{1}+\mu _{2}Q_{2}\right) \right) ^{2}-\left( \mu _{1}\mu _{2}\varrho
-\left( \mu _{2}P_{1}-\mu _{1}P_{2}\right) \right) ^{2}\right] \right\}
\colon  \label{13}
\end{equation}%
where $\lambda \equiv \mu _{1}^{2}+\mu _{2}^{2}$, using $\colon \exp \left(
-a_{1}^{\dag }a_{1}-a_{2}^{\dag }a_{2}\right) \colon $ $=|00\rangle \langle
00|$, and $C$ is the normalized constant that can be determined by
\begin{equation*}
\left\langle q_{cm},\varrho \right\vert \left. q_{cm}^{\prime },\varrho
^{\prime }\right\rangle =\delta \left( \varrho ^{\prime }-\varrho \right)
\delta \left( q_{cm}^{\prime }-q_{cm}\right) ,\qquad {}\left( \mu _{1}=\mu
_{2}\right)
\end{equation*}%
up to a random phase, and this can be evaluated easily by insert
overcompleteness of coherent state $\int d^{2}z\left\vert z\right\rangle
\left\langle z\right\vert =\pi \,\,$, where $\left\vert z\right\rangle
=D\left( z\right) \left\vert 0\right\rangle $, and $D\left( z\right) =\exp %
\left[ za^{\dagger }-z^{\ast }a\right] $ is displacement operator, $%
\left\langle z^{\prime }\right\vert \left. z\right\rangle =\exp \left[
z^{\prime \ast }z-\left( \left\vert z\right\vert ^{2}+\left\vert z^{\prime
}\right\vert ^{2}\right) /2\right] $, so ($z_{i}=x_{i}+\mathrm{i}y_{i}$,
where $x_{i}$ and $y_{i}$ are real and image part of $z_{i}$)
\begin{eqnarray}
&&\int \frac{d^{2}z_{1}d^{2}z_{2}}{\pi ^{2}}\left\langle q_{cm},\varrho
\right\vert \left. z_{1}z_{2}\right\rangle \left\langle
z_{1}z_{2}\right\vert \left. q_{cm}^{\prime },\varrho ^{\prime }\right\rangle
\notag \\
&=&\left\vert C\right\vert ^{2}\int \frac{d^{2}z_{1}d^{2}z_{2}}{\pi ^{2}}%
\frac{\mu _{1}\mu _{2}}{\pi \lambda }\exp \left\{ \frac{1}{\lambda }\left[
\sqrt{2}\mu _{1}\left( q_{cm}^{\prime }+\mathrm{i}\varrho ^{\prime }\mu
_{2}^{2}\right) z_{1}^{\ast }+\sqrt{2}\mu _{2}\left( q_{cm}^{\prime }-%
\mathrm{i}\varrho ^{\prime }\mu _{1}^{2}\right) z_{2}^{\ast }\right. \right.
\\
&&\left. \left. -\frac{1}{2}(\mu _{1}^{2}-\mu _{2}^{2})\left( z_{1}^{\ast
}z_{1}^{\ast }-z_{2}^{\ast }z_{2}^{\ast }\right) -2\mu _{1}\mu
_{2}z_{1}^{\ast }z_{2}^{\ast }-\frac{1}{2}(q_{c}^{\prime 2}+\left( \mu
_{1}\mu _{2}\varrho ^{\prime }\right) ^{2})\right] \right\}  \notag \\
&&\exp \left[ -z_{1}z_{1}^{\ast }-z_{2}z_{2}^{\ast }\right] \exp \left\{
\frac{1}{\lambda }\left[ \sqrt{2}\mu _{1}\left( q_{cm}-\mathrm{i}\varrho \mu
_{2}^{2}\right) z_{1}+\sqrt{2}\mu _{2}\left( q_{cm}+\mathrm{i}\varrho \mu
_{1}^{2}\right) z_{2}\right. \right.  \notag \\
&&\left. \left. -\frac{1}{2}(\mu _{1}^{2}-\mu _{2}^{2})\left(
z_{1}z_{1}-z_{2}z_{2}\right) -2\mu _{1}\mu _{2}z_{1}z_{2}-\frac{1}{2}%
(q_{c}^{2}+\left( \mu _{1}\mu _{2}\varrho \right) ^{2})\right] \right\}
\notag \\
&=&\left\vert C\right\vert ^{2}\int \frac{dx_{2}dy_{2}}{2\pi ^{2}}\exp \left[
-i\sqrt{2}\mu _{2}\left( \varrho ^{\prime }-\varrho \right) x_{2}-i\sqrt{2}%
\left( q_{cm}^{\prime }-q_{cm}\right) \frac{y_{2}}{\mu _{2}}+C(\varrho
^{\prime },\varrho ,q_{cm}^{\prime },q_{cm})\right]  \notag \\
&=&\left\vert C\right\vert ^{2}\delta \left( \varrho ^{\prime }-\varrho
\right) \delta \left( q_{cm}^{\prime }-q_{cm}\right) \exp \left[ C(\varrho
^{\prime },\varrho ,q_{cm}^{\prime },q_{cm})\right]  \label{inner_product}
\end{eqnarray}%
where
\begin{eqnarray*}
C(\varrho ^{\prime },\varrho ,q_{cm}^{\prime },q_{cm}) &=&-\frac{2i\mu
_{2}^{2}\left[ \mu _{1}^{2}\left( \varrho ^{\prime }+\varrho \right) \left(
q_{cm}^{\prime }-q_{cm}\right) -\mu _{2}^{2}\left( \varrho ^{\prime
}-\varrho \right) \left( q_{cm}^{\prime }+q_{cm}\right) \right] }{\left[
4\mu _{2}^{2}\left( \mu _{1}-i\mu _{2}\right) \left( \mu _{1}+i\mu
_{2}\right) \right] } \\
&&-\frac{\left( \mu _{1}^{2}+\mu _{2}^{2}\right) \left[ \mu _{2}^{4}\left(
\varrho ^{\prime }-\varrho \right) ^{2}+\left( q_{cm}^{\prime
}-q_{cm}\right) ^{2}\right] }{\left[ 4\mu _{2}^{2}\left( \mu _{1}-i\mu
_{2}\right) \left( \mu _{1}+i\mu _{2}\right) \right] }
\end{eqnarray*}

When $\lambda =\mu _{1}=\mu _{2}=1/2$, equation (\ref{inner_product}) writes$%
\bigskip $%
\begin{equation*}
\int \frac{d^{2}z_{1}d^{2}z_{2}}{\pi ^{2}}\left\langle q_{cm},\varrho
\right\vert \left. z_{1}z_{2}\right\rangle \left\langle
z_{1}z_{2}\right\vert \left. q_{cm}^{\prime },\varrho ^{\prime
}\right\rangle =\left\vert C\right\vert ^{2}\delta \left( \varrho ^{\prime
}-\varrho \right) \delta \left( q_{cm}^{\prime }-q_{cm}\right)
\end{equation*}
so, we can select $C=1$, and the decomposition yields
\begin{eqnarray}
\left\vert q_{cm},\varrho \right\rangle &=&\sqrt{\frac{\mu _{1}\mu _{2}}{\pi
\lambda }}\exp \left\{ \frac{1}{\lambda }\left[ \sqrt{2}\mu _{1}\left(
q_{cm}+\mathrm{i}\varrho \mu _{2}^{2}\right) a_{1}^{\dagger }+\sqrt{2}\mu
_{2}\left( q_{cm}-\mathrm{i}\varrho \mu _{1}^{2}\right) a_{2}^{\dagger
}\right. \right.  \notag \\
&&\left. \left. -\frac{1}{2}(\mu _{1}^{2}-\mu _{2}^{2})\left( a_{1}^{\dagger
}a_{1}^{\dagger }-a_{2}^{\dagger }a_{2}^{\dagger }\right) -2\mu _{1}\mu
_{2}a_{1}^{\dagger }a_{2}^{\dagger }-\frac{1}{2}(q_{c}^{2}+\left( \mu
_{1}\mu _{2}\varrho \right) ^{2})\right] \right\} \left\vert 00\right\rangle
.  \label{qp-bi}
\end{eqnarray}

If we set $q_{cm}=\sqrt{\lambda }\xi _{q}$, $\varrho =\sqrt{\lambda }\xi
_{p} $, and $\xi =$ $\xi _{q}+\mathrm{i}\xi _{p}$, then equation (\ref{qp-bi}%
) can be rewritten as
\begin{eqnarray}
\left\vert \xi \right\rangle &=&\frac{\sqrt{\mu _{1}\mu _{2}}}{\lambda }\exp
\left\{ -\frac{1}{2}\left\vert \xi \right\vert ^{2}+\frac{1}{\sqrt{2\lambda }%
}\left[ \xi +\left( \mu _{1}-\mu _{2}\right) \xi ^{\ast }\right]
a_{1}^{\dagger }+\frac{1}{\sqrt{2\lambda }}\left[ \xi ^{\ast }-\left( \mu
_{1}-\mu _{2}\right) \xi \right] a_{2}^{\dagger }\right.  \notag \\
&&\left. -\frac{1}{2\lambda }\left( \mu _{1}-\mu _{2}\right) \left(
a_{1}^{\dagger 2}-a_{2}^{\dagger 2}\right) -4\mu _{1}\mu _{2}a_{1}^{\dagger
}a_{2}^{\dagger }\right\} \left\vert 00\right\rangle \text{.}
\end{eqnarray}

When $\lambda =\mu _{1}=\mu _{2}=1/2$,
\begin{equation*}
\left\vert \xi \right\rangle =\exp \left[ -\frac{1}{2}\left\vert \xi
\right\vert ^{2}+\xi a_{1}^{\dagger }+\xi ^{\ast }a_{2}^{\dagger
}-a_{1}^{\dagger }a_{2}^{\dagger }\right] \left\vert 00\right\rangle
\end{equation*}
This $\left\vert \xi \right\rangle $ is the conjugate state of $\left\vert
\eta \right\rangle $ in (\ref{state-yita}). Using the IWOP technique we have%
\begin{eqnarray}
\int \frac{d^{2}\xi }{\pi }\left\vert \xi \right\rangle \left\langle \xi
\right\vert &=&\iint \mathrm{d}q_{cm}\mathrm{d}\varrho \frac{\mu _{1}\mu _{2}%
}{\pi \lambda }\colon \exp \left\{ \frac{1}{\lambda }\left[ -\left(
q_{cm}-\left( \mu _{1}Q_{1}+\mu _{2}Q_{2}\right) \right) ^{2}-\left( \mu
_{1}\mu _{2}\varrho -\left( \mu _{2}P_{1}-\mu _{1}P_{2}\right) \right) ^{2}%
\right] \right\} \colon  \notag \\
&=&1\text{.}
\end{eqnarray}%
This is a convenient approach for finding bipartite entangled state.{} and
its eigen-equations write ($\xi =$ $\xi _{q}+\mathrm{i}\xi _{p}$),
\begin{eqnarray}
Q_{cm}\left\vert \xi \right\rangle &=&\sqrt{\lambda }\xi _{q}\left\vert \xi
\right\rangle \text{,}  \notag \\
P_{r}\left\vert \xi \right\rangle &=&\sqrt{\lambda }\xi _{p}\left\vert \xi
\right\rangle \text{,}
\end{eqnarray}

\section{Deriving the common entangled eigenstates of tripartite's
center-of-mass coordinate and mass-weighted relative momenta}

\label{tripartite}

Having experienced how to find the common eigenstates of bipartite's
center-of-mass coordinate and mass-weighted relative momentum via the Weyl
correspondence and the IWOP technique, we now search for the common
eigenvector of the tripartite. Noticing that the three compatible operators:
the center-of-mass coordinate $\mu _{1}Q_{1}+\mu _{2}Q_{2}+\mu _{3}Q_{3}$,
and $P_{1}/\mu _{1}-P_{2}/\mu _{2}$, $P_{1}/\mu _{1}-P_{3}/\mu _{3}$, which
are mass-weighted relative momentums, where $\mu _{i}$ is the relative of
mass defined $\mu _{i}=m_{i}/M,$ and $m_{i}$ is the mass of the $i$-th
particle, $M=m_{1}+m_{2}+m_{3}$ is the total mass of all three particles, so
$\mu _{1}+\mu _{2}+\mu _{3}=1$. Since the three operators are complete and
compatible with each other, there is a representation spaned by their common
eigenvector $\left\vert q,\varrho _{2},\varrho _{3}\right\rangle $, which
satisfies the following eigen-equations
\begin{subequations}
\begin{eqnarray}
\left( \mu _{1}Q_{1}+\mu _{2}Q_{2}+\mu _{3}Q_{3}\right) \left\vert q,\varrho
_{2},\varrho _{3}\right\rangle &=&q\left\vert q,\varrho _{2},\varrho
_{3}\right\rangle . \\
\left( \frac{P_{1}}{\mu _{1}}-\frac{P_{2}}{\mu _{2}}\right) \left\vert
q,\varrho _{2},\varrho _{3}\right\rangle &=&\varrho _{2}\left\vert q,\varrho
_{2},\varrho _{3}\right\rangle , \\
\left( \frac{P_{1}}{\mu _{1}}-\frac{P_{3}}{\mu _{3}}\right) \left\vert
q,\varrho _{2},\varrho _{3}\right\rangle &=&\varrho _{3}\left\vert q,\varrho
_{2},\varrho _{3}\right\rangle ,
\end{eqnarray}%
In similar to (\ref{10})
\end{subequations}
\begin{eqnarray}
\left\vert q,\varrho _{2},\varrho _{3}\right\rangle \left\langle q,\varrho
_{2},\varrho _{3}\right\vert &\Longrightarrow &\delta \left\{ \left[ \varrho
_{2}-\left( \frac{p_{1}}{\mu _{1}}-\frac{p_{2}}{\mu _{2}}\right) \right]
\right\} \delta \left[ \varrho _{3}-\left( \frac{p_{1}}{\mu _{1}}-\frac{p_{3}%
}{\mu _{3}}\right) \right]  \notag \\
&&\times \delta \left[ q-\left( \mu _{1}q_{1}+\mu _{2}q_{2}+\mu
_{3}q_{3}\right) \right] .
\end{eqnarray}%
Accordingly, we have
\begin{eqnarray}
&&\left\vert q,\varrho _{2},\varrho _{3}\right\rangle \left\langle q,\varrho
_{2},\varrho _{3}\right\vert  \notag \\
&\Longrightarrow &\idotsint \prod\limits_{i=1}^{3}\mathrm{d}q_{i}\mathrm{d}%
p_{i}\delta \left[ \varrho _{2}-\left( \frac{p_{1}}{\mu _{1}}-\frac{p_{2}}{%
\mu _{2}}\right) \right] \delta \left[ \varrho _{3}-\left( \frac{p_{1}}{\mu
_{1}}-\frac{p_{3}}{\mu _{3}}\right) \right]  \notag \\
&&\times \delta \left[ q-\left( \mu _{1}q_{1}+\mu _{2}q_{2}+\mu
_{3}q_{3}\right) \right] \Delta _{1}\left( q_{1},p_{1}\right) \Delta
_{2}\left( q_{2},p_{2}\right) \Delta _{3}\left( q_{3},p_{3}\right)  \notag \\
&=&\idotsint \prod\limits_{i=1}^{3}\mathrm{d}q_{i}\mathrm{d}p_{i}\delta %
\left[ \varrho _{2}-\left( \frac{p_{1}}{\mu _{1}}-\frac{p_{2}}{\mu _{2}}%
\right) \right] \delta \left[ \varrho _{3}-\left( \frac{p_{1}}{\mu _{1}}-%
\frac{p_{3}}{\mu _{3}}\right) \right]  \notag \\
&&\times \delta \left[ q-\left( \mu _{1}q_{1}+\mu _{2}q_{2}+\mu
_{3}q_{3}\right) \right] \frac{1}{\pi ^{3}}\colon \exp \left\{
\sum\limits_{i=1}^{3}\left[ -\left( q_{i}-Q_{i}\right) ^{2}-\left(
p_{i}-P_{i}\right) ^{2}\right] \right\} \colon  \notag \\
&=&\iiint \frac{\mu _{2}}{\pi ^{3}}\mathrm{d}p_{1}\mathrm{d}q_{1}\mathrm{d}%
q_{2}\exp \left\{ -\left( q_{1}-Q_{1}\right) ^{2}-\left( p_{1}-P_{1}\right)
^{2}-\left( q_{2}-Q_{2}\right) ^{2}\right.  \notag \\
&&\left. -\left( \frac{q-\mu _{1}q_{1}-\mu _{2}q_{2}}{\mu _{3}}-Q_{3}\right)
^{2}-\left( \frac{\mu _{2}}{\mu _{1}}p_{1}-\mu _{2}\varrho _{2}-P_{2}\right)
^{2}-\left( \frac{\mu _{3}}{\mu _{1}}p_{1}-\mu _{3}\varrho _{3}-P_{3}\right)
^{2}\right\}  \notag \\
&=&\pi ^{-3/2}\frac{\mu _{1}\mu _{2}\mu _{3}}{\lambda }\exp \left\{ -\frac{1%
}{\lambda }\left[ \left[ \mu _{1}\mu _{2}\varrho _{2}-\left( \mu
_{2}P_{1}-\mu _{1}P_{2}\right) \right] ^{2}+\left[ \mu _{1}\mu _{3}\varrho
_{3}-\left( \mu _{3}P_{1}-\mu _{1}P_{3}\right) \right] ^{2}\right. \right.
\notag \\
&&\left. \left. +\left[ \mu _{2}\mu _{3}\left( \varrho _{2}-\varrho
_{3}\right) +\left( \mu _{3}P_{2}-\mu _{2}P_{3}\right) \right] ^{2}-\left(
q-\sum\limits_{i=1}^{3}\mu _{i}Q_{i}\right) ^{2}\right] \right\}
\end{eqnarray}
where $\lambda =\sum_{i=1}^{3}\mu _{i}^{2}$,
\begin{eqnarray}
&&\left\vert q,\varrho _{2},\varrho _{3}\right\rangle \left\langle q,\varrho
_{2},\varrho _{3}\right\vert  \notag \\
&=&\left\vert C\right\vert ^{2}\pi ^{-3/2}\frac{\mu _{1}\mu _{2}\mu _{3}}{%
\lambda }\exp \left\{ -\frac{1}{\lambda }\left[ \left[ \mu _{1}\mu
_{2}\varrho _{2}-\left( \mu _{2}P_{1}-\mu _{1}P_{2}\right) \right] ^{2}+%
\left[ \mu _{1}\mu _{3}\varrho _{3}-\left( \mu _{3}P_{1}-\mu
_{1}P_{3}\right) \right] ^{2}\right. \right. \\
&&\left. \left. +\left[ \mu _{2}\mu _{3}\left( \varrho _{2}-\varrho
_{3}\right) +\left( \mu _{3}P_{2}-\mu _{2}P_{3}\right) \right] ^{2}-\left(
q-\sum\limits_{i=1}^{3}\mu _{i}Q_{i}\right) ^{2}\right] \right\}  \label{18}
\end{eqnarray}

\bigskip the normailization constant C is determined by
\begin{equation*}
\left\langle q,\varrho _{2},\varrho _{3}\right\vert \left. q^{\prime
},\varrho _{2}^{\prime },\varrho _{3}^{\prime }\right\rangle =\delta \left(
q^{\prime }-q\right) \delta \left( \varrho _{2}^{\prime }-\varrho
_{2}\right) \delta \left( \varrho _{3}^{\prime }-\varrho _{3}\right) ,\qquad{%
}(\mu _{1}=\mu _{2}=\mu _{3})
\end{equation*}

\ \bigskip Similar procedure to the bipartite situation works out C=1, So
splitting the right hand side of (\ref{18}) as the form $f\left( a_{i}^{\dag
}\right) |000\rangle \langle 000|f^{\dagger }\left( a_{i}^{\dagger }\right)
, $ where%
\begin{equation}
\colon \exp \left( -\sum\limits_{i=1}^{3}a_{i}^{\dag }a_{i}\right) \colon
=|000\rangle \langle 000|,  \label{19}
\end{equation}

After the decomposition, we can have \cite{r22}%
\begin{eqnarray}
\left\vert q,\varrho _{2},\varrho _{3}\right\rangle &=&\pi ^{-3/4}\sqrt{%
\frac{\mu _{1}\mu _{2}\mu _{3}}{\lambda }}\exp \left\{ A+\frac{\sqrt{2}q}{%
\lambda }\sum_{i=1}^{3}\mu _{i}a_{i}^{\dagger }+\right.  \notag \\
&&+\frac{\mathrm{i}\sqrt{2}\mu _{2}\varrho _{2}}{\lambda }\left[ \mu _{1}\mu
_{2}a_{1}^{\dagger }-\left( \mu _{1}^{2}+\mu _{3}^{2}\right) a_{2}^{\dagger
}+\mu _{2}\mu _{3}a_{3}^{\dagger }\right]  \notag \\
&&\left. +\frac{\mathrm{i}\sqrt{2}\mu _{3}\varrho _{3}}{\lambda }\left[ \mu
_{1}\mu _{3}a_{1}^{\dagger }+\mu _{2}\mu _{3}a_{2}^{\dagger }-\left( \mu
_{1}^{2}+\mu _{3}^{2}\right) a_{3}^{\dagger }\right] +S^{\dagger }\right\}
|000  \label{tri-state}
\end{eqnarray}
Where
\begin{eqnarray*}
A &=&-\frac{q^{2}}{2\lambda }-\frac{1}{2\lambda }\left[ -2\mu _{2}^{2}\mu
_{3}^{2}\varrho _{2}\varrho _{3}+\left( \mu _{1}^{2}+\mu _{3}^{2}\right) \mu
_{2}^{2}\varrho _{2}^{2}+\left( \mu _{1}^{2}+\mu _{2}^{2}\right) \mu
_{3}^{2}\varrho _{3}^{2}\right] \\
S &=&-\frac{1}{\lambda }\sum_{i,j=1}^{3}\left( \mu _{i}\mu _{j}a_{i}a_{j}-%
\frac{\lambda }{2}\delta _{ij}\right)
\end{eqnarray*}

When $\lambda =\mu _{1}=\mu _{2}=\mu _{3}=1/3$, the equation (\ref{tri-state}%
) can be rewritten to

\begin{eqnarray}
\left\vert q,\varrho _{2},\varrho _{3}\right\rangle &=&\frac{1}{3\pi ^{3/4}}%
\exp \left\{ -\frac{3}{2}q^{2}+\sqrt{2}q\left( a_{1}^{\dagger
2}+a_{2}^{\dagger 2}+a_{3}^{\dagger 2}\right) -\frac{2}{3}\left(
a_{1}^{\dagger }a_{2}^{\dagger }+a_{1}^{\dagger }a_{3}^{\dagger
}+a_{2}^{\dagger }a_{3}^{\dagger }\right) \right. \\
&&+\frac{\sqrt{2}\mathrm{i}}{9}\left[ \left( \varrho _{2}+\varrho
_{3}\right) a_{1}^{\dagger }+\left( 2\varrho _{2}-\varrho _{3}\right)
a_{2}^{\dagger }+\left( 2\varrho _{3}-\varrho _{2}\right) a_{3}^{\dagger }%
\right] \\
&&\left. -\frac{1}{27}\left( \varrho _{2}^{2}+\varrho _{3}^{2}-\varrho
_{2}\varrho _{3}\right) \right\} |000\rangle
\end{eqnarray}

Using the IWOP technique and Eq. (\ref{18}) we have%
\begin{eqnarray}
&&\iiint_{-\infty }^{\infty }\mathrm{d}q\mathrm{d}\varrho _{2}\mathrm{d}%
\varrho _{3}\left\vert q,\varrho _{2},\varrho _{3}\right\rangle \left\langle
q,\varrho _{2},\varrho _{3}\right\vert  \notag \\
&=&\iiint_{-\infty }^{\infty }\mathrm{d}q\mathrm{d}\varrho _{2}\mathrm{d}%
\varrho _{3}\pi ^{-3/2}\frac{\mu _{1}\mu _{2}\mu _{3}}{\lambda }\exp \left\{
-\frac{1}{\lambda }\left[ \left[ \mu _{1}\mu _{2}\varrho _{2}-\left( \mu
_{2}P_{1}-\mu _{1}P_{2}\right) \right] ^{2}+\left[ \mu _{1}\mu _{3}\varrho
_{3}-\left( \mu _{3}P_{1}-\mu _{1}P_{3}\right) \right] ^{2}\right. \right.
\notag \\
&&\left. \left. +\left[ \mu _{2}\mu _{3}\left( \varrho _{2}-\varrho
_{3}\right) +\left( \mu _{3}P_{2}-\mu _{2}P_{3}\right) \right] ^{2}-\left(
q-\sum\limits_{i=1}^{3}\mu _{i}Q_{i}\right) ^{2}\right] \right\}  \notag \\
&=&\colon \exp \left( \mathbf{0}\right) \colon =1.
\end{eqnarray}

\bigskip The Weyl correspondence approach is a very direct way to find the
tripartite entangled state $\left\vert q,\varrho _{2},\varrho
_{3}\right\rangle $, which make up a complete set.

\section{Deriving the entangled eigenstate of multipartite's center-of-mass
coordinate and mass-weighted relative momenta}

\label{multipartite}

Enlightened by the former method, we now search for the mass-dependent
multipartite entangled system via the Weyl correspondence and the IWOP
technique, we introduce commute operators: the center-of-mass coordinate $%
\sum_{i=1}^{n}\mu _{i}Q_{i}$, and $P_{j}/\mu _{j}-P_{i}/\mu _{i}$, ($%
i,j=1,2,\cdots ,n$), the mass-weighted \ and relative momentums, where $\mu
_{i}$ is the relative mass defined $\mu _{i}=m_{i}/M,$ $m_{i}$ is the mass
of the $i$-th particle, $M=\sum_{i=1}^{n}m_{i}$ is the total mass of all
particles, and $\sum_{i=1}^{n}\mu _{i}=1$. Since these operators are commute
with each other, there will be common eigenvectors, denoted as $\left\vert
q,\varrho _{2},\varrho _{3,}\cdots \varrho _{n}\right\rangle $ for them,
which satisfies the following eigen-equations
\begin{subequations}
\begin{eqnarray}
\sum_{i=1}^{n}\mu _{i}Q_{i}\left\vert q,\varrho _{2},\varrho _{3,}\cdots
\varrho _{n}\right\rangle &=&q\left\vert q,\varrho _{2},\varrho _{3,}\cdots
\varrho _{n}\right\rangle .  \label{eigen-q-multi} \\
\left( \frac{P_{1}}{\mu _{1}}-\frac{P_{i}}{\mu _{i}}\right) \left\vert
q,\varrho _{2},\varrho _{3,}\cdots \varrho _{n}\right\rangle &=&\varrho
_{i}\left\vert q,\varrho _{2},\varrho _{3,}\cdots \varrho _{n}\right\rangle
,\qquad i=2,\cdots ,n  \label{eigen-p-multi}
\end{eqnarray}%
From the above discussion and the eigen-equations of $\left\vert q,\varrho
_{2},\varrho _{3,}\cdots \varrho _{n}\right\rangle $, we immediately write
down the classical Weyl correspondence of the projector $\left\vert
q,\varrho _{2},\varrho _{3,}\cdots \varrho _{n}\right\rangle \left\langle
q,\varrho _{2},\varrho _{3,}\cdots \varrho _{n}\right\vert $, we have
\end{subequations}
\begin{equation}
\left\vert q,\varrho _{2},\varrho _{3,}\cdots \varrho _{n}\right\rangle
\left\langle q,\varrho _{2},\varrho _{3,}\cdots \varrho _{n}\right\vert
\Longrightarrow \delta \left( q-\sum\limits_{i=1}^{n}\mu _{i}q_{i}\right)
\prod\limits_{j=2}^{n}\delta \left[ \varrho _{j}-\left( \frac{p_{1}}{\mu _{1}%
}-\frac{p_{j}}{\mu _{j}}\right) \right] \prod\limits_{i=1}^{n}\Delta
_{i}\left( q_{i},p_{i}\right) ,
\end{equation}%
that is
\begin{eqnarray}
&&\left\vert q,\varrho _{2},\varrho _{3,}\cdots \varrho _{n}\right\rangle
\left\langle q,\varrho _{2},\varrho _{3,}\cdots \varrho _{n}\right\vert
\notag \\
&\Longrightarrow &\idotsint \prod\limits_{i=1}^{n}\mathrm{d}q_{i}\mathrm{d}%
p_{i}\delta \left( q-\sum\limits_{i=1}^{n}\mu _{i}q_{i}\right)
\prod\limits_{j=2}^{n}\delta \left[ \varrho _{j}-\left( \frac{p_{1}}{\mu _{1}%
}-\frac{p_{j}}{\mu _{j}}\right) \right] \prod\limits_{i=1}^{n}\Delta
_{i}\left( q_{i},p_{i}\right)  \notag \\
&=&\frac{1}{\pi ^{n}}\idotsint \prod\limits_{i=1}^{n}\mathrm{d}q_{i}\mathrm{d%
}p_{i}\delta \left( q-\sum\limits_{i=1}^{n}\mu _{i}q_{i}\right)
\prod\limits_{j=2}^{n}\delta \left[ \varrho _{j}-\left( \frac{p_{1}}{\mu _{1}%
}-\frac{p_{j}}{\mu _{j}}\right) \right]  \notag \\
&&\times \colon \exp \left\{ \sum\limits_{i=1}^{n}\left[ -\left(
q_{i}-Q_{i}\right) ^{2}-\left( p_{i}-P_{i}\right) ^{2}\right] \right\} \colon
\notag \\
&=&\frac{1}{\pi ^{n}}\frac{\prod_{i=1}^{n}\mu _{i}}{\mu _{1}\mu _{n}}%
\idotsint \mathrm{d}p_{1}\prod\limits_{i=1}^{n-1}\mathrm{d}q_{i}\colon \exp %
\left[ -\sum\limits_{i=1}^{n-1}\left( q_{i}-Q_{i}\right) ^{2}-\left( \frac{%
q-\sum_{i=1}^{n-1}\mu _{i}q_{i}}{\mu _{n}}-Q_{n}\right) ^{2}\right.  \notag
\\
&&\left. -\left( p_{1}-P_{1}\right) ^{2}-\sum\limits_{i=2}^{n}\left(
-P_{i}\right) ^{2}\right] \colon  \notag \\
&=&\frac{1}{\pi ^{n}}\frac{\prod_{i=1}^{n}\mu _{i}}{\mu _{1}\mu _{n}}\int
\mathrm{d}p_{1}\colon \exp \left[ -\left( p_{1}-P_{1}\right)
^{2}-\sum\limits_{i=2}^{n}\left( \frac{\mu _{i}}{\mu _{1}}p_{1}-\mu
_{i}\varrho _{i}-P_{i}\right) ^{2}\right] \colon  \notag \\
&&\times \idotsint \prod\limits_{i=1}^{n-1}\mathrm{d}q_{i}\colon \exp \left[
-\left( \frac{q}{\mu _{n}}-\sum\limits_{i=1}^{n-1}\frac{\mu _{i}q_{i}}{\mu
_{n}}-Q_{n}\right) ^{2}-\sum\limits_{i=1}^{n-1}\left( q_{i}-Q_{i}\right) ^{2}%
\right] \colon  \label{multi-partite}
\end{eqnarray}

where
\begin{equation}
\int \mathrm{d}p_{1}\colon \exp \left[ -\left( p_{1}-P_{1}\right)
^{2}-\sum\limits_{i=2}^{n}\left( \frac{\mu _{i}}{\mu _{1}}p_{1}-\mu
_{i}\varrho _{i}-P_{i}\right) ^{2}\right] \colon =\mu _{1}\sqrt{\frac{\pi }{%
\lambda }}N  \label{26}
\end{equation}%
and $\lambda =\sum_{i=1}^{n}\mu _{i}^{2}$, and we introduced $\varrho _{1}=0$
for simplification of $N$
\begin{eqnarray*}
N &=&\colon \exp \left\{ -\frac{1}{\lambda }\left[ \sum_{j=2}^{n}\left[ \mu
_{1}\mu _{j}\varrho _{j}-\left( \mu _{j}P_{1}-\mu _{1}P_{j}\right) \right]
^{2}+\sum_{j<k=2}^{n}\left[ \mu _{j}\mu _{k}\left( \varrho _{j}-\varrho
_{k}\right) +\left( \mu _{k}P_{j}-\mu _{j}P_{k}\right) \right] ^{2}\right]
\right\} \colon \\
&=&\colon \exp \left\{ -\frac{1}{2\lambda }\left[ \sum_{j,k=1}^{n}\left[ \mu
_{j}\mu _{k}\left( \varrho _{j}-\varrho _{k}\right) +\left( \mu
_{k}P_{j}-\mu _{j}P_{k}\right) \right] ^{2}\right] \right\} \colon
\end{eqnarray*}

To integrate over the remaining part of equation (\ref{multi-partite}), we
resort to the following mathematical formula%
\begin{equation}
\int ...\int_{-\infty }^{\infty }d^{n}\chi \exp [-\widetilde{\mathbf{\chi }}%
\mathbf{B\chi }+\widetilde{\mathbf{\chi }}\mathbf{\upsilon }]=\sqrt{\frac{%
\pi ^{n}}{\det \mathbf{B}}}\exp [\frac{1}{4}\widetilde{\mathbf{\upsilon }}%
\mathbf{B}^{-1}\mathbf{\upsilon }],  \label{formula}
\end{equation}%
where $\mathbf{B}$ is a symmetric positive-definite invertible covariant
tensor of rank $n$, $\widetilde{\mathbf{\chi }}=\left( \chi _{1},\chi
_{2},\chi _{3},\cdots ,\chi _{n}\right) $ is transpose of $\mathbf{\chi }$,
and so we can obtain%
\begin{eqnarray}
&&\idotsint \prod\limits_{i=1}^{n-1}\mathrm{d}q_{i}\colon \exp \left[
-\left( \frac{q}{\mu _{n}}-\sum\limits_{i=1}^{n-1}\frac{\mu _{i}q_{i}}{\mu
_{n}}-Q_{n}\right) ^{2}-\sum\limits_{i=1}^{n-1}\left( q_{i}-Q_{i}\right) ^{2}%
\right] \colon  \notag \\
&=&\idotsint \prod\limits_{i=1}^{n-1}\mathrm{d}q_{i}\colon \exp \left[
-\left( q_{1},q_{2},\cdots ,q_{n-1}\right) \mathbf{B}\left(
\begin{array}{c}
q_{1} \\
q_{2} \\
\vdots \\
q_{n-1}%
\end{array}%
\right) +\left( q_{1},q_{2},\cdots ,q_{n-1}\right) \left(
\begin{array}{c}
\upsilon _{1} \\
\upsilon _{2} \\
\vdots \\
\upsilon _{n-1}%
\end{array}%
\right) -\mathit{const}\right] \colon  \notag \\
\end{eqnarray}%
and
\begin{subequations}
\begin{eqnarray}
\mathbf{B} &=&\left( B_{ij}\right) _{\left( n-1\right) \times \left(
n-1\right) }, \\
B_{ij} &=&\frac{\mu _{i}\mu _{j}}{\mu _{n}^{2}}+\delta _{ij}, \\
\upsilon _{i} &=&2\frac{\mu _{i}}{\mu _{n}^{2}}q+2Q_{i}-2\frac{\mu _{i}}{\mu
_{n}}Q_{n}, \\
\det \mathbf{B} &=&1+\sum_{i=1}^{n-1}\left( \frac{\mu _{i}}{\mu _{n}}\right)
^{2}=\frac{\lambda }{\mu _{n}^{2}}, \\
\mathit{const} &=&\left( \frac{q}{\mu _{n}}-Q_{n}\right)
^{2}+\sum\limits_{i=1}^{n-1}Q_{i}^{2}
\end{eqnarray}

So the inverse of the matrix $\mathbf{B}$ is

\begin{eqnarray}
\mathbf{B}^{-1} &=&\left( B_{ij}^{\prime }\right) _{\left( n-1\right) \times
\left( n-1\right) } \\
B_{ij}^{\prime } &=&\frac{1}{\det \mathbf{B}}\left( \delta _{ij}\det \mathbf{%
B}-\frac{\mu _{i}\mu _{j}}{\mu _{n}^{2}}\right)
\end{eqnarray}

\bigskip and using the formula (\ref{formula}), we have

\end{subequations}
\begin{eqnarray}
&&\idotsint \prod\limits_{i=1}^{n-1}\mathrm{d}q_{i}\colon \exp \left[
-\left( \frac{q}{\mu _{n}}-\sum\limits_{i=1}^{n-1}\frac{\mu _{i}q_{i}}{\mu
_{n}}-Q_{n}\right) ^{2}-\sum\limits_{i=1}^{n-1}\left( q_{i}-Q_{i}\right) ^{2}%
\right] \colon  \notag \\
&=&\frac{\pi ^{\left( n-1\right) /2}}{\sqrt{\det \mathbf{B}}}\colon \exp %
\left[ -\frac{1}{\lambda }\left( q-\sum\limits_{i=1}^{n}\mu _{i}Q_{i}\right)
^{2}\right] \colon ,  \label{27}
\end{eqnarray}

\bigskip Substituting (\ref{26}) and (\ref{27}) into (\ref{multi-partite}),
we have

\begin{equation}
\left\vert q,\varrho _{2},\varrho _{3,}\cdots \varrho _{n}\right\rangle
\left\langle q,\varrho _{2},\varrho _{3,}\cdots \varrho _{n}\right\vert =\pi
^{-n/2}\frac{\prod_{i=1}^{n}\mu _{i}}{\lambda }\left\vert C\right\vert
^{2}\colon N\exp \left[ -\frac{1}{\lambda }\left( q-\sum\limits_{i=1}^{n}\mu
_{i}Q_{i}\right) ^{2}\right] \colon .  \label{multi-bra-ket}
\end{equation}

\bigskip Similarly, the $C$ constant can be select to unit, since
\begin{equation*}
\left\langle q,\varrho _{2},\varrho _{3,}\cdots \varrho _{n}\right.
\left\vert q^{\prime },\varrho _{2}^{\prime },\varrho _{3,}^{\prime }\cdots
\varrho _{n}^{\prime }\right\rangle =\delta \left( q^{\prime }-q\right)
\prod_{i=2}^{n}\delta \left( \varrho _{i}^{\prime }-,\varrho _{i,}\right)
,\qquad {}\left( \mu _{i}=\lambda \right)
\end{equation*}%
and decomposing the right hand side of (\ref{multi-bra-ket}) as the form $%
f\left( a_{i}^{\dag }\right) |00\cdots 0\rangle \langle 00\cdots
0|f^{\dagger }\left( a_{i}^{\dagger }\right) $, where%
\begin{equation}
\colon \exp \left( -\sum\limits_{i=1}^{n}a_{i}^{\dag }a_{i}\right) \colon
=|00\cdots 0\rangle \langle 00\cdots 0|,
\end{equation}%
we get
\begin{equation}
\left\vert q,\varrho _{2},\varrho _{3,}\cdots \varrho _{n}\right\rangle =\pi
^{-n/4}\sqrt{\frac{\prod_{i=1}^{n}\mu _{i}}{\lambda }}\exp \left\{ \frac{1}{%
\lambda }\left[ \frac{M}{2}+\sqrt{2}\sum\limits_{i=1}^{n}A_{i}a_{i}^{\dag
}+\sum\limits_{i,j=1}^{n}K_{ij}a_{i}^{\dag }a_{j}^{\dag }\right] \right\}
|00\cdots 0\rangle ,  \label{multi-state}
\end{equation}%
where (please attention $\varrho _{1}=0$).
\begin{subequations}
\begin{eqnarray}
A_{i} &=&\mu _{i}q-\mathrm{i}\sum_{j=1}^{n}\left[ \mu _{i}\mu _{j}^{2}\left(
\varrho _{i}-\varrho _{j}\right) \right] \\
K_{ij} &=&-\mu _{i}\mu _{j}+\frac{\delta _{ij}}{2}\lambda \\
M &=&-q^{2}-\sum_{k=1}^{n}\mu _{1}^{2}\mu _{k}^{2}\varrho _{k}^{2}-\frac{1}{2%
}\sum_{k,l=2}^{n}\mu _{l}^{2}\mu _{k}^{2}\left( \varrho _{k}-\varrho
_{l}\right) ^{2}  \notag \\
&=&-q^{2}-\frac{1}{2}\sum_{i,j=1}^{n}\left[ \mu _{i}\mu _{j}\left( \varrho
_{i}-\varrho _{j}\right) \right] ^{2}
\end{eqnarray}

By this concise approach, we find the multipartite EPR entangled state \cite%
{r20,r22}. When $\lambda =\mu _{i}=1/n$ ($i=1,2,\cdots ,n$), the equation (%
\ref{tri-state}) can be rewritten to \bigskip \cite{r21}

\end{subequations}
\begin{eqnarray}
\left\vert q,\varrho _{2},\varrho _{3,}\cdots \varrho _{n}\right\rangle
&=&\pi ^{-n/4}n^{(1-n)/2}\exp \left\{ \sum_{j,k=1}^{n}\left[ \left( \frac{1}{%
2}a_{j}^{\dagger }a_{k}^{\dagger }-\frac{q^{2}}{2}+\sqrt{2}qa_{j}^{\dagger
}\right) \delta _{jk}\right. \right.  \notag \\
&&\left. \left. -\frac{1}{n}a_{j}^{\dagger }a_{k}^{\dagger }-\frac{\sqrt{2}%
\mathrm{i}}{n^{2}}\left( \varrho _{j}-\varrho _{k}\right) a_{j}^{\dagger }-%
\frac{1}{4n^{3}}\left( \varrho _{j}-\varrho _{k}\right) ^{2}\right] \right\}
|00\cdots 0\rangle ,
\end{eqnarray}

It is quite straightforward to demonstrate that equation (\ref{multi-state})
is the eigenvector of the center-of-mass coordinate $\sum_{i=1}^{n}\mu
_{i}Q_{i}$ and mass-weighted relative momentums $P_{1}/\mu _{1}-P_{i}/\mu
_{i}$, ($i=1,2,\cdots ,n$) with eigenvalue $q$, $\varrho _{i}$ respective,
i.e. (\ref{eigen-q-multi}) and (\ref{eigen-p-multi}), and its completeness
writes
\begin{eqnarray}
&&\idotsint_{-\infty }^{\infty }\mathrm{d}q\prod\limits_{i=2}^{n}\mathrm{d}%
\varrho _{i}\left\vert q,\varrho _{2},\varrho _{3,}\cdots \varrho
_{n}\right\rangle \left\langle q,\varrho _{2},\varrho _{3,}\cdots \varrho
_{n}\right\vert  \notag \\
&=&\idotsint_{-\infty }^{\infty }\mathrm{d}q\prod\limits_{i=2}^{n}\mathrm{d}%
\varrho _{i}\pi ^{-n/2}\frac{\prod_{i=1}^{n}\mu _{i}}{\lambda }\colon N\exp %
\left[ -\frac{1}{\lambda }\left( q-\sum\limits_{i=1}^{n}\mu _{i}Q_{i}\right)
^{2}\right] \colon  \notag \\
&=&\colon \exp \left( \mathbf{0}\right) \colon =1.
\end{eqnarray}

\bigskip So, we have derived the multipartite EPR entangled representation
of multi-mode via the Weyl correspondence approach.

\section{Conclusion}

\label{conclusion}

Due to the IWOP technique and the Weyl correspondence (Weyl quantization
scheme) we have presented a new concise approach for obtaining the Fock
representation of multi-partite entangled states of continuum variables. \
This is one available approach for finding many new quantum mechanical
representations which may enrich Dirac's representation and transformation
theory. In this paper, we employed this new concise approach from Weyl
correspondence to derive the entended Fan-Klauder entangled state
representation to multipartite case representation. Our derivation itself
demonstrate the effective and efficient of the approach for search new
representations.

\section{\protect\bigskip Acknowledgements}

This work was supported by the President Foundation of Chinese Academy of
Science and the National Natural Science Foundation of China under grant
10475056.

\end{document}